# Synthesis, Structure and Properties of Boron and Nitrogen Doped Graphene

L. S. Panchakarla[1], K. S. Subrahmanyam[1], S. K. Saha[2], A. Govindaraj[1], H. R. Krishnamurthy[2], U. V. Waghmare[3*], and C. N. R. Rao[1*]

[1]*Chemistry and Physics of Materials Unit, New Chemistry Unit and CSIR Centre of Excellence in Chemistry, Jawaharlal Nehru Centre for Advanced Scientific Research, Jakkur P.O., Bangalore 560064, India*

[2]*Department of Physics, Indian Institute of Science, Bangalore 560012, India*

[3]*Theoretical Sciences Unit, Jawaharlal Nehru Centre for Advanced Scientific Research, Bangalore 560064, India*

[*]**email:** cnrrao@jncasr.ac.in, Fax: (+91) 80-2208 2760, waghmare@jncasr.ac.in

Graphene has emerged as an exciting material today because of the novel properties associated with its two-dimensional structure.[1,2] Single-layer graphene is a one-atom thick sheet of carbon atoms densely packed into a two-dimensional honeycomb lattice. It is the mother of all graphitic forms of carbon including zero-dimensional fullerenes and one-dimensional carbon nanotubes.[1] The remarkable feature of graphene is that it is a Dirac solid, with the electron energy being linearly dependent on the wave vector near the vertices of the hexagonal Brillouin zone. It exhibits room-temperature fractional quantum Hall effect[3] and ambipolar electric field effect along with ballistic conduction of charge carriers.[4] It has been reported recently that a top-gated single layer-graphene transistor is able to reach electron or hole doping levels of upto $5 \times 10^{13}$ cm$^{-2}$. The doping effects are ideally monitored by Raman spectroscopy.[5-10] Thus, the G band in the Raman spectrum stiffens for both electron and hole doping and the ratio of the intensities of the 2D and G band varies sensitively with doping. Molecular charge-transfer induced by electron-donor and -acceptor molecules also gives rise to significant changes in the electronic structure of few-layer graphenes, as evidenced by changes in the Raman spectrum.[6,7] Charge-transfer by donor and acceptor molecules soften and stiffen the G band respectively. The difference between electrochemical doping and doping through molecular charge transfer is noteworthy. It is of fundamental interest to investigate how these effects compare with the effects of doping graphene by substitution with boron and nitrogen and to understand dopant-induced perturbations on the properties of graphene. Secondly, opening of the band-gap in graphene is essential to facilitate its applications in electronics and graphene bilayers[11] are an attractive option for this. With this motivation, we have prepared, for the first time, B- and N-doped graphene bilayer samples employing different strategies and investigated their structure and properties. We have also carried out first-principles DFT calculations to understand the effect of substitutional doping on the structure of graphene as well as its electronic and vibrational properties.

To prepare B-and N-doped graphenes, we have exploited our recent finding that arc discharge between carbon electrodes in a hydrogen atmosphere yields graphenes (HG) with two to three layers. We have prepared B-doped graphene (BG) by two methods involving arc discharge of graphite electrodes in the presence $H_2+B_2H_6$ (BG1) and by carrying out arc discharge using a boron-stuffed graphite electrode (BG2). We have prepared nitrogen-doped graphene (NG) by carrying out arc discharge in the presence of $H_2$+pyridine (NG1) or $H_2$+ammonia (NG2). We have also carried out the transformation of nanodiamond in the presence of pyridine (NG3) to obtain N-doped graphene. All the doped samples were characterized by variety of physical methods.

X-ray photoelectron spectroscopic (XPS) analysis showed that the BG1 and BG2 contained 1.2 and 3.1 at% of boron respectively, while the electron energy loss spectroscopy (EELS) data showed the content of boron in these samples to be 1.0 and 2.4 at% respectively. In Fig. 1a, we show typical core-level XPS data of BG2 along with the elemental mapping by EELS. XPS data show NG1, NG2 and NG3 to contain 0.6, 0.9 and 1.4 at% of nitrogen respectively. In Fig. 1b, we have shown XPS data of the NG2 sample along with EELS elemental mapping. The asymmetric shape of the N 1s peak indicates the existence of at least two components. On deconvolution, we find peaks at 398.3 eV and 400 eV, the first one being characteristic of pyridinic nitrogen ($sp^2$ hybridization) and the second of nitrogen in the graphene sheets.[12,13] Analysis of the (002) reflections in the X-ray diffraction (XRD) patterns showed that the B-and N-doped samples contained 2-3 layers on an average. This is also confirmed by the TEM images (see Fig. 2 for typical TEM images). Atomic force microscopy images also showed the presence of 2-3 layers in the BG and NG samples, with occasional presence of single layers. TGA shows that the B- and N- doped samples undergo combustion at a slightly lower temperature than the parent graphene (580 $^o$C).

We have examined the Raman spectra of all the BG and NG samples in comparison with the spectrum of the pure graphene sample (HG), prepared by the $H_2$-discharge method. We show typical spectra in Fig. 3. The Raman spectrum [with 632.8 nm excitation] of these samples shows three main features in 1000-3000 cm$^{-1}$ region, the G band (~ 1570 cm$^{-1}$), the defect-related D band[14] (~ 1320 cm$^{-1}$), and the 2D band(~ 2640 cm$^{-1}$). It is noteworthy that the G band stiffens both with boron and nitrogen doping. This is similar to what happens with electrochemical doping.[5] The shift in the case of BG2, with a higher B-content, is larger than with BG1. In the case of N-doped graphene, NG3, with the highest N content, shows the largest shift. The intensity of the D band is higher with respect to that of the G band in all the doped samples. On doping, the relative intensity of the 2D band generally decreases with respect to the G band. We have calculated the in-plane crystallite sizes ($L_a$) of the undoped as well as doped graphene samples by following formula,[15]

$$L_a(nm) = (2.4 \times 10^{-10}) \lambda^4 (I_D/I_G)^{-1}.$$

Here $\lambda$ is wavelength used for Raman measurements and $I_D$, $I_G$ respectively intensity of D and G band. The crystallite size of the HG, BG1, BG2, NG1, NG2 and NG3 samples are respectively 64, 30, 26, 43, 41 and 19 nm. Doped graphene samples show low in crystallite sizes compared to undoped graphene samples. We also find that the BG and NG samples exhibit much lower electrical resistivity than that of undoped graphene.

We have considered two configurations of doped bi-layer graphene in our simulation (4x4 supercell), where the substituted (3.125%) atoms (B or N at a time) in the two layers are (a) close to and (b) far from each other. We find that configurations with dopant atoms separated by larger distance from each other are more favourable energetically. While the configuration with widely separated N atoms is lower in energy by 25 meV than the one with N at nearest sites in the two planes, this energy difference for B-substitution is rather small (4 meV). This implies that homogeneous B-substitution may be easier than N-substitution. The origin of this difference can be traced to the structure: B-C bond is about 0.5% longer than the C-C bond while N-C bond is about the same as C-C bond in length, resulting in significant relaxation of the structure of B-doped bilayer dominating its energetics. The interplanar separation reduces by almost 2.7% in B-doped bilayer graphene while it remains almost unchanged in N-doped bilayers. We estimate the energy of formation of doped graphene from graphene and dopant atoms in the gaseous form to be 5.6 and 8.0 eV/atom for B and N doping respectively, suggesting that synthesis of B- and N-doped graphene should be quite possible.

Our calculations reveal that the linearity in the dispersion of the electronic bands within 1eV of the Fermi-energy is almost unchanged with B- and N-doping (see Fig 4)! This means that the doped graphenes have the potential to exhibit the interesting properties of pristine graphenes. Fermi-energy which is at the apex of the conical band-structure near the **K**-point of Brillouin zone of graphene, is shifted by -0.65 eV and 0.59 eV in the case of 2 at% B and N-substitutions. These shifts are of -1.0 and 0.9 eV in 3.125 at% B- and N-doped bilayers respectively. This results in p-type and n-type semi-conducting behaviour of graphene (see Fig. 4), as expected. We note that much (96%) of the shift in Fermi energy arises essentially from the substitution with dopants, and the remaining 4% arises from the lattice relaxation. This is also reflected in the almost symmetric shift in Fermi energy with B- and N-substitution, which were seen to be associated with large and small structural relaxation respectively. It is indeed encouraging because this means that effects of doping should scale reasonably well to lower concentrations.

As Raman spectroscopy is ideally suited for the characterization of doped graphene[5-10], it is important to know how it correlates with the concentration of carriers or dopants. The shift in the G-band frequency measured by Raman spectroscopy has many physical contributions and we use our calculations to uncover the relative magnitudes of these. Unexpected purely on the basis of the lengths of B-C and N-C bonds, in the context of carbon nano-tubes, Yang *et al*[16] and Coville *et al*[17,18] had reported an interesting finding that the G band shifts are in the same direction. We demonstrate (see Table I) that indeed the shifts in vibrational frequencies of graphene with B and N-doping have opposite signs if one takes into account only the changes in bond-length obtained at fixed lattice constant. Relaxation or the change of lattice constant is highly asymmetric: the lattice constant increases by 0.32% with 2% boron substitution in single layer graphene, while it decreases very slightly with N-substitution, resulting in a large decrease and a slight increase in frequency respectively. With these two mechanisms alone, the shift in G-band frequency with either B or N substitution is negative, contrary to our measured trends. After adding the dynamic corrections[19-21], these frequency shifts become positive, in agreement with experimental observations reported here. Very similar changes in the G-band of bilayer graphene (see Table I) are estimated from our calculations with B and N substitution, assuming the same amount of dynamic corrections as in the mono-layer. This assumption is reasonable for shifts in Fermi energy above 0.39 eV and less than 1 eV (the second band is populated by electrons or holes, as seen in Figure 3), because the density of states of the bilayer matches with that of the monolayer and hence the dynamic corrections are expected to be similar (see Figs. 3 and 4 in Ref. 18). Thus, the comparison between experiment and calculations shown here can be used to conclude that dynamic corrections are important in understanding the changes in the phonon frequencies of graphene and of bilayers arising from the introduction of carriers. In fact, the overall neutrality of substitutionally doped graphene allows a very clean and accurate calculation and further testing of the role of dynamic corrections arising from doping to the G-band in graphene.

Our observation on the non-segregating tendency or the homogeneity in the distribution of boron implies that the disorder or the number of possible configurations will increase with the concentration of dopant atoms and result in more prominent peaks of the D-band, as evidenced in the measured Raman spectra reported here [see Fig-3]. Secondly, the changes in frequency of the Raman 2D-band can be understood from our results for the phonons at **K**+Δ**K** accessible in the spectrum of Γ-point phonons, determined for the 4x4 supercell. We find that the shifts in the frequency of phonons at **K**+Δ**K** are comparable to the observed values (see Table I, lower panel) and much weaker in magnitude than the ones at **K**, re-affirming the strong Kohn-anomaly in the phonons at the **K**-point. We note that the eigen-modes of the G-band and D-band (see Figure 5) also change with doping by developing features in the atomic displacements localized near the dopant atoms, providing a physical picture why these modes could be used to characterize the nature of dopant atom and its effects on the electronic structure of graphene or of bilayer-graphene.

Having demonstrated that different routes are possible for synthesis of B and N-doped graphene, and based on a consistent agreement between our experiments and calculations, and complementary information derived on the electronic structure, we conclude that B- and N-doped graphene can be synthesized to exhibit p- and n- type semiconducting electronic properties that can be systematically tuned with concentration of B and N, and can be characterized with Raman spectroscopy of the G- and D-bands. Realization of such p- and n-type conducting bilayers should be usable in a variety of devices similar to the ones based on semiconductors.

*Experimental*

*Synthesis of boron and nitrogen doped graphene:* One set of boron doped graphene samples (BG1) was prepared by carrying out arc discharge of graphite electrodes in the presence of hydrogen, helium and diborane ($B_2H_6$). $B_2H_6$ vapor was generated by the addition of $BF_3$-diethyl etherate to sodium borohydride in tetraglyme. $B_2H_6$ vapor was carried to the arc chamber by passing hydrogen (200 torr) through $B_2H_6$ generator and subsequently by passing He (500 torr). Second set of boron doped samples (BG2) was prepared by carrying out arc discharge using a boron-stuffed graphite electrode (3 at% boron) in the presence of $H_2$ (200 torr) and He (500 Torr).

One set of nitrogen doped graphene samples (NG1) was prepared by carrying out arc discharge of graphite electrodes in the presence of $H_2$, He and pyridine vapors. Pyridine vapour was carried to the arc chamber by passing hydrogen (200 torr) through pyridine bubbler and subsequently by passing He (500 torr). Second set of nitrogen doped samples (NG2) was prepared by carrying out arc discharge of graphite electrodes in the presence of $H_2$ (200 torr), He (200 torr) and NH3 (300 torr). Transformation of nanodiamond was also carried out in the presence of He and Pyridine vapour at 1650 $^o$C to obtain N-doped graphene (NG3). We have also carried out the transformation of nanodiamond in the presence of pyridine (NG3) to obtain N-doped graphene.

*Sample characterization:* X-ray diffraction (XRD) patterns of the samples were recorded in the θ-2 θ Bragg-Bretano geometry with a siemens D5005 diffractometer using Cu $K_α$ (λ =0.151418nm) radiation. Raman spectra were recorded with LabRAM HR high resolution

Raman spectrometer (Horiba Jobin Yvon) using He-Ne Laser (λ=632.8 nm). Transmission electron microscope (TEM) images were obtained with a JEOL JEM 3010 instrument. Atomic force microscope (AFM) measurements were performed using NanoMan. X-ray photoelectron spectroscopy (XPS) was recorded using a VG scientific ESCA LAB V spectrometer. EELS were recorded with a transsmision electron microscope (FEI, TECNAI F30) equiped with an energy filter for EELS operating at 200kV. Thermogravimetric analysis was carried out using Mettler Toledo TGA 850 instrument.

*Computational Methods:* Our first-principles calculations are based on ultra-soft pseudopotentials and density functional theory as implemented in Quantum Espresso[22]. The electron exchange correlation energy has been approximated with the PBE[23] form of the generalized gradient approximation functional for single-layer graphene and the PZ[24] form of the local density approximation functional for bi-layer graphene, as the latter describes the binding between the two graphene planes more realistically. Doping with B and N was simulated by replacing one carbon atom in a 5x5 (4x4) supercell, amounting to 2 (3.125) atomic % substitution in the graphene monolayer (bilayer). In each case, the structure was fully relaxed to the lowest energy. Kohn-Sham wave functions were expressed in the plane-wave basis with an energy cut off of 40 Ry and integrations over the Brillouin zones were sampled with a 9x9x1 (6x6x1 for bilayer) mesh of uniformly-spaced **k**-points. Vibrational properties were determined using density functional theory linear response formalism within the adiabatic approximation, and further dynamic corrections were obtained using the time-dependent perturbation theory[20] and the linear form of electronic band dispersion at the Fermi level. Our methodology and calculational parameters have been tested for convergence before.[25,26]

**Acknowledgements**

L.S.P. and K.S.S. gratefully acknowledge CSIR, New Delhi, for a senior research fellowship and S.K.S. acknowledges financial support from the JNCASR. We would also like to acknowledge use of central computing facility from the Centre for Computational Materials Science at JNCASR.

Figure captions

Fig. 1. (a) C 1s and B 1s XPS signals of B-doped graphene (BG2), EELS elemental mapping of C and B of BG2. (b) C 1s and N 1s XPS signals of N-doped graphene (NG2), EELS elemental mapping of C and N of NG2.

Fig. 2. TEM images of (a) B-doped graphene (BG2), (b) N-doped graphene (NG1). Calculated scanning tunnelling microscopy (STM) images of (c) B- and (d) N-doped bilayers. B and N doping results in depletion or addition of electronic charge on carbon atoms on the sublattice of the substituted dopant, as evident in weaker green (B) and blue colors (N) respectively.

Fig. 3. Raman spectra of undoped (HG) and doped (BG and NG) graphene samples.

Fig. 4. Electronic structure of substitutionally doped graphene monolayer (a) and (b), and bilayers (c) and (d), with boron and nitrogen respectively. $a_s$ is the in-plane lattice constant of the supercell. It is remarkable that doping results in a small gap opened up in mono-layer graphene and a weak quadratic dispersion in bilayer. This is ideal for applications of graphene in electronic devices.

Fig. 5. Atomic displacements of the G-band and D-band modes calculated (a) and (d) for pristine bilayers, (b) and (e) for B doped bilayers, and (c) and (f) for N-doped bilayers. The extended modes evolve to exhibit localized features upon doping: boron atom being lighter displaces more than carbon and displacement of N-atom is much weaker than of carbon. Frequencies given here are before adding dynamical corrections.

Table I: Various contributions to the shifts in phonon frequencies of graphene and bilayers resulting from B or N substitution. "Fixed lattice" means internally relaxed structure keeping the lattice constant the same as of undoped system; "Relaxed lattice" means the lattice constant is also relaxed for doped configurations. Dynamic corrections are obtained using formalism in Refs. 19 and 20.

| G-band | Fixed lattice | Relaxed lattice | With dynamic correction |
|---|---|---|---|
| Pristine monolayer graphene | 1570.7 | 1570.7 | 1570.7 |
| 2% Boron doped monolayer | 1579.4 | 1561.2 | 1584.6 |
| 2% Nitrogen doped monolayer | 1546.5 | 1553.4 | 1574.4 |
| Pristine bilayer graphene | 1603.8 | 1603.8 | 1603.8 |
| 3.125% Boron doped bilayer | 1609.7 | 1582.4 | 1617.8 |
| 3.125% Nitrogen doped bilayer | 1566.1 | 1577.7 | 1609.9 |
| D-bands | Frequency (cm$^{-1}$) | D-band shift (cm$^{-1}$) | 2D-band shift (cm$^{-1}$) |
| Pristine bilayer graphene | 1404.1 | 0.0 | 0.0 |
| 3.125% Boron doped bilayer | 1387.2 | -16.9 | -33.2 |
| 3.125% Nitrogen doped bilayer | 1400.0 | -4.1 | -8.2 |

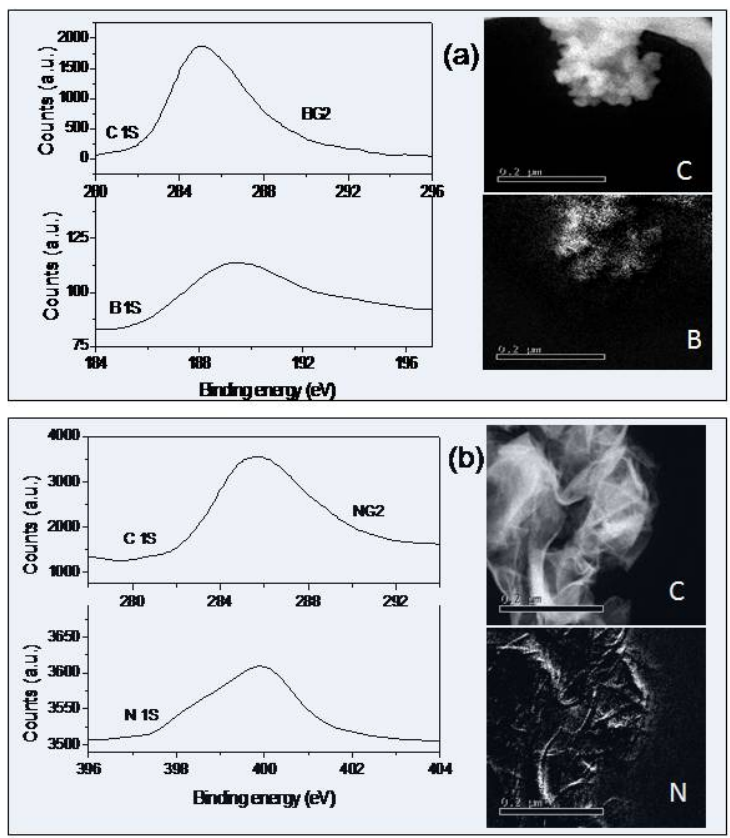

Fig. 1

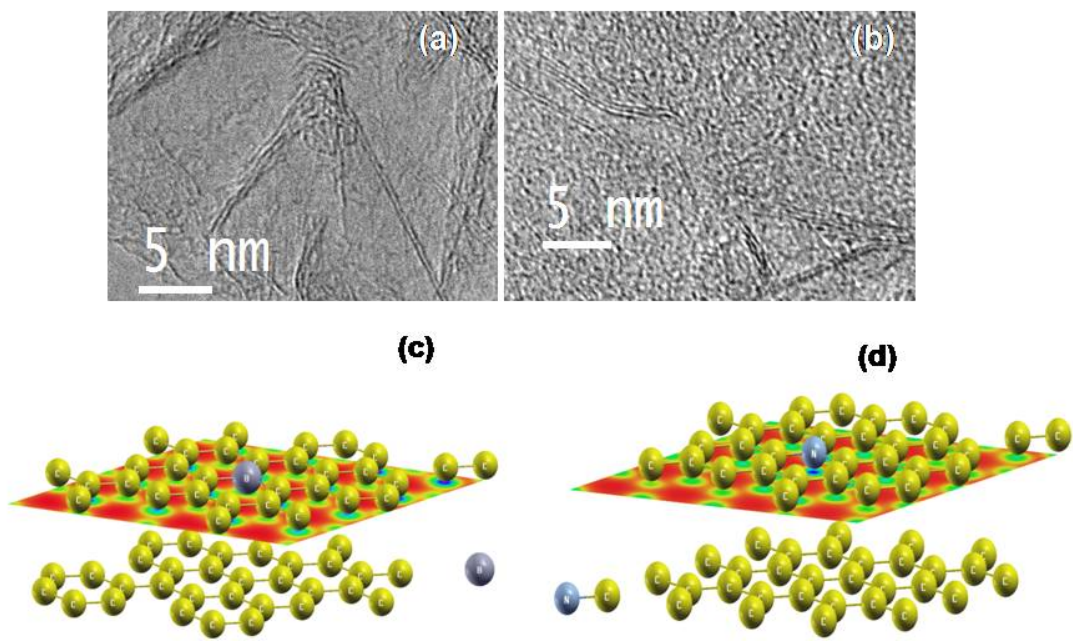

Fig. 2

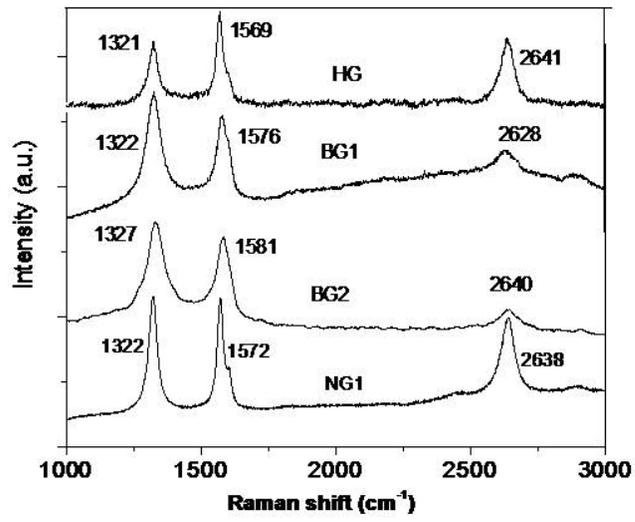

Fig. 3

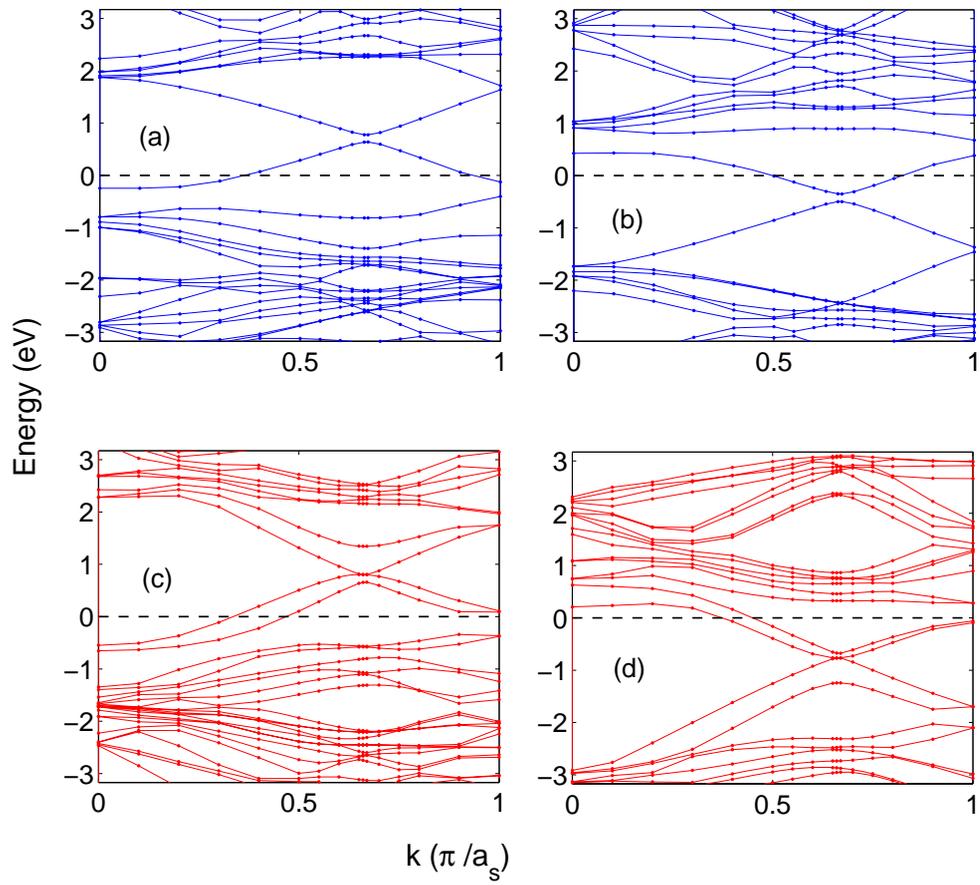

Fig. 4

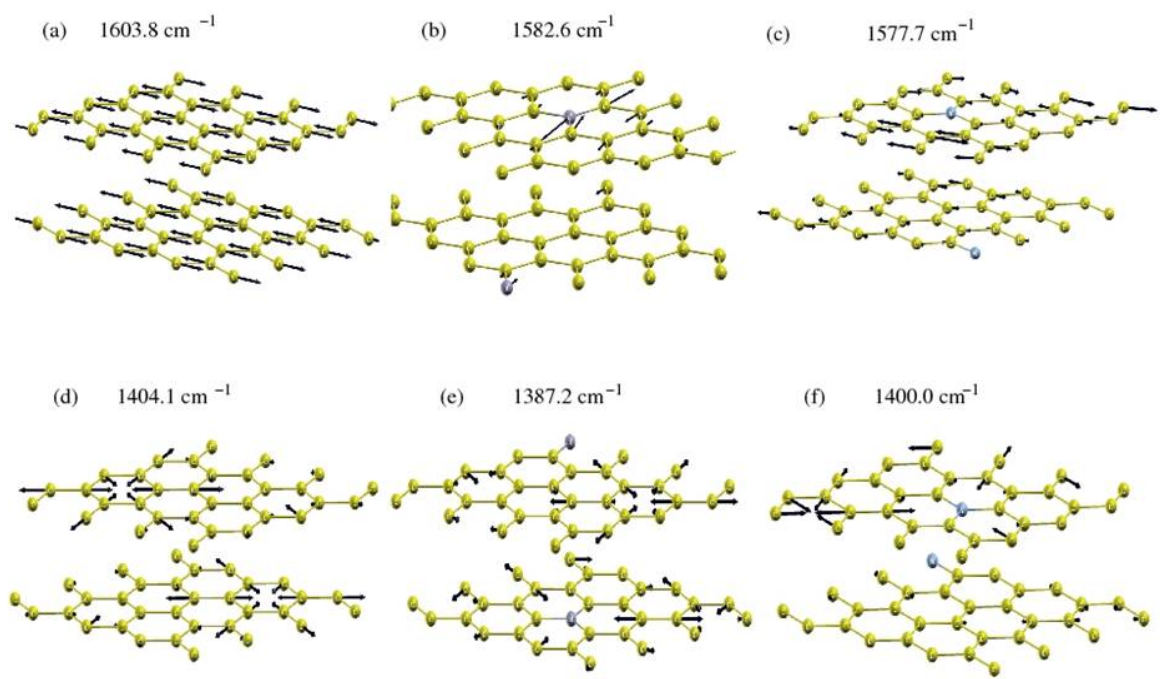

Fig. 5

**Graphical abstract:**

We present the structure and properties of boron and nitrogen doped graphenes, obtained by more than one method involving arc discharge between carbon electrods or by the transformation of nano-diamond in an appropriate gaseous atmosphere. Using a combination of experiment and first-principles theory, we demonstrate systematic changes in the carrier-concentration and electronic structure of graphenes with B/N-doping, accompanied by stiffening of the G-band and intensification of the defect related D-band in the Raman spectra.

**Graphical figure**

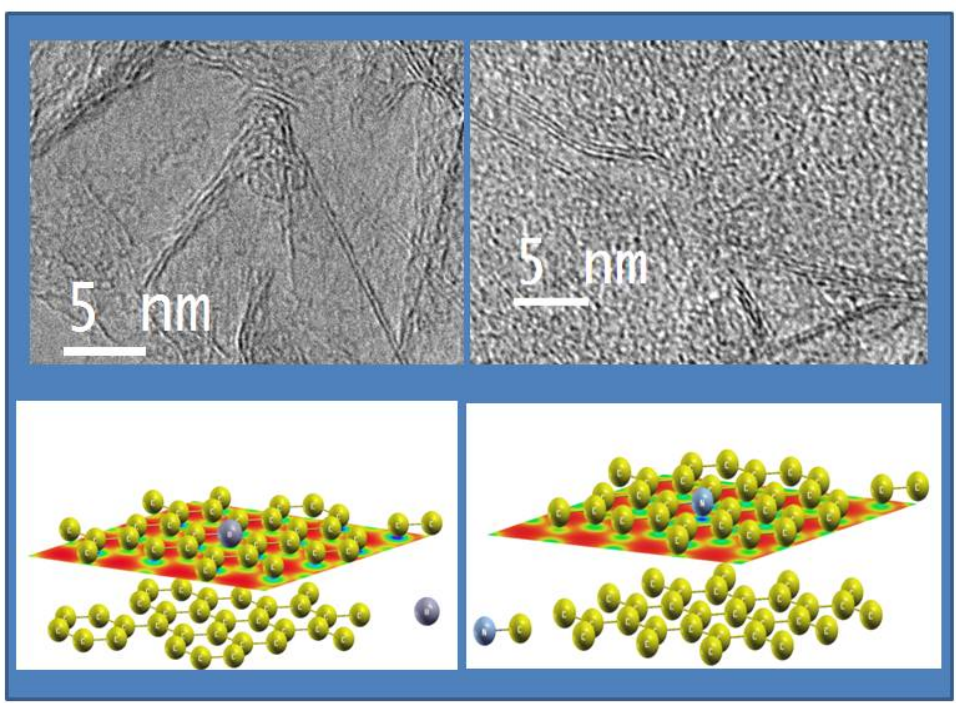